# Low-Cost Traffic Sensing System Based on LoRaWAN for Urban Areas

Hannaneh Barahouei Pasandi[1], Asma Haqiqat[2], Azin Moradbeikie[2,3], Ahmad Keshavarz[2], Habib Rostami[2], Sara Paiva[3], Sérgio Ivan Lopes[3,4,5]

[1]Virginia Commonwealth University, Richmond, VA 23284, USA
[2]Persian Gulf University, Shahid Mahini Street, Bushehr, Iran
[3]ADiT-Lab, Instituto Politecnico de Viana do Castelo, 4900-348 Viana do Castelo, Portugal
[4]CiTin—Centro de Interface Tecnologico Industrial, 4970-786 Arcos de Valdevez, Portugal
[5]IT—Instituto de Telecomunicacoes, Campus Universitario de Santiago, 3810-193 Aveiro, Portugal

## ABSTRACT

The advent of Low Power Wide Area Networks (LPWAN) has enabled the feasibility of wireless sensor networks for environmental traffic sensing across urban areas. In this study, we explore the usage of LoRaWAN end nodes as traffic sensing sensors to offer a practical traffic management solution. The monitored Received Signal Strength Indicator (RSSI) factor is reported and used in the gateways to assess the traffic of the environment. Our technique utilizes LoRaWAN as a long-range communication technology to provide a large-scale system. In this work, we present a method of using LoRaWAN devices to estimate traffic flows. LoRaWAN end devices then transmit their packets to different gateways. Their RSSI will be affected by the number of cars present on the roadway. We used SVM and clustering methods to classify the approximate number of cars present. This paper details our experiences with the design and real implementation of this system across an area that stretches for miles in urban scenarios. We continuously measured and reported RSSI at different gateways for weeks. Results have shown that if a LoRaWAN end node is placed in an optimal position, up to 96% of correct environment traffic level detection can be obtained. Additionally, we share the lessons learned from such a deployment for traffic sensing.

## 1 INTRODUCTION

According to the United Nations, nowadays, 55% of the world's population lives in cities, which is predicted to increase to 68% by 2050 [1]. LoRa promises city-scale IoT deployments for smart city applications due to its long-distance communication capabilities [35]. With such rapid growth, successful urban planning and management rely on adopting efficient and smart applications such as traffic sensing that are able to provide important information for essential decision-making systems. In recent years, the conventional wireless communication infrastructure is shifted to not only be used for communication but to provide and support novel and emerging low-cost technologies. Efficient smart services city management has reached a broad appeal ranging from smart healthcare to traffic management and pedestrian monitoring, just to name a few. In today's world, smart traffic management emphasizes sensing, analysis, control, and communication with ground transportation to provide safe and light traffic management.

The main purpose of traffic management systems is to provide useful information including Road Information Services (RIS), traffic congestion monitoring, real-time route planning, and road surface monitoring [22]. To enable such services, the smart traffic management system is highly dependent on adopting an effective traffic sensing method. Recent advances in traffic sensing methods enhance the traffic decision making. They collects sensed data to obtain contextualized information from the surrounding environment and then transmit it to a remote server for additional processing and analytics to improve decision-making [21, 23, 29–31].

Environmental data gathering typically relies upon a network of IoT devices that not only collects data but also performs its aggregation and transformation to generate contextualized sensing information. However, the issue of adopting IoT technologies for collecting data presents a fundamental challenge that arises from the fact that the data sources need to be low-cost and have low power consumption. However, using specialized sensors for this purpose, the deployment will become costly at a large scale. To make the matter even worse, the implemented network that provides the connection between end nodes and the server should handle long-range coverage. There are multiple attempts to use radio signals for determining the environmental features, for avoiding sensor implementation costs. In recent years, using Low Power Wide Area Networks (LPWAN) has been proposed as a candidate solution to provide a communication network between IoT end devices and the application server. In this paper we focus on exploring the usage of conventional LoRaWAN end nodes as traffic sensors to offer a practical traffic sensing and control solution. We demonstrate the feasibility of our approach. If a LoRaWAN end node is placed in an optimal position, up to 96% of correct environment traffic level detection can be obtained. Moreover, source code has been made available online as open access[1]. LPWAN communication technologies can be useful for traffic sensing and traffic control in outdoor settings. Other technologies have been the subject of prior research for applications similar to traffic sensing [24]. We argue that such a design is impractical for large deployments. The access to the signal phase information that is necessary for these designs to work requires specific hardware. Few Wi-Fi devices offer this data to application developers. The term "LPWAN" refers to a variety of technologies, including LTE-M, NB-IoT, Sigfox, and LoRaWAN, that enable long-distance communication between end nodes at a cheap cost and with little power consumption [26]. LoRaWAN is a good candidate for traffic sensing since it has advantages over competitors in terms of deployment, range coverage, battery life, cost-effectiveness, and latency. Therefore, in this study, we investigate the usage of LoRaWAN end nodes as traffic-sensing sensors to provide a practical traffic control strategy. In the proposed method, the Received Signal Strength Indicator (RSSI) factor is measured and reported at gateways to estimate the traffic size in the surrounding area to develop a large-scale system by utilizing LoRaWAN.

## 2 RELATED WORKS

In recent years, wireless signals have been used in numerous sensing applications. One of the main applications is traffic sensing for traffic management systems. These technologies offer traffic awareness for managing experience time and jamming. On the one hand, increasing the urbanization rate poses a challenge to designing a cost-effective system to detect crowded areas. On the other hand, with the wide deployment of IoT devices across cities, solving the traffic problem is feasible based on previous studies [2]. Currently, proposed traffic management methods for transportation systems utilize various strategies to control traffic and congestion in different areas. These methods can be divided into different categories based on six attributes: (a) Model Type, (b) Sensing Technologies (c) Data Gathering Techniques (d) Selected Road Infrastructure (e) Traffic Management, and (f) Result Verification Approach [4]. In the following, we describe those lines of research that falls into sensing technologies approaches.

In the following, we divide the literature based on Traffic data collection using different technologies and approaches.
**RFID and Sensors.** Authors in [5] proposed an architecture to integrate data collected from IoT devices, RFID, wireless sensor network (WSN), GPS, cloud computing, agent, and other advanced technologies to store, manage and supervise traffic information. Besides proposing a complex approach that requires many resources, their proposed architecture demands a large number of RFIDs and sensors which makes it less practical and more time-consuming. May research efforts in the literature, focus on the use of equipped vehicles with different sensors for data collection [6] (more examples). Such methods not only could violate the owner's privacy, but most importantly they are also not cost-efficient.
**Unmanned Aerial Vehicles.** The emergence of Unmanned Aerial Vehicles (UAVs) has enabled the possibility of providing a wide range of smart city management including smart traffic management. Different countries initiated numerous traffic management projects based on UAVs [11]. Authors in [13] proposed a decentralized UAV traffic management protocol that ensures high levels of integrity, availability, and confidentiality operations. For this purpose, the Authors exploit the blockchain and smart contract technologies. However, The proposed protocol is not a scalable, practical solution in all locations with different scenarios.
**Wireless Sensor Networks (WSNs).** WSNs are another approach to managing Urban traffic. They are capable of detecting the related parameters of traffic. flows [25]. Despite such positive capabilities, the main drawbacks of WSN-based methods are being expensive and highly dependent on weather and/or lighting conditions that may require special maintenance that making them even less unsuitable for large-area deployment.
**Signal-based methods** Such methods are less sensitive to light or weather compared to WSNs which could largely impact their usage for large area coverage and improved accuracy. Such features make them more suitable for traffic monitoring applications.

In this study, we proposed to use LoRaWAN technology. We measured and monitored the received RSSI of end nodes

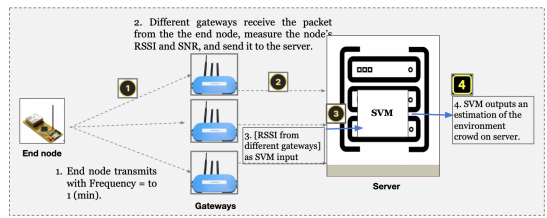

**Figure 1: Proposed method for spatial data collection in LoRaWAN testbed.**

---

across the city. The end nodes are repurposed to help us provide an estimation of environment traffic sensing by measuring valid metrics. In addition to being considered cost-effective, our approach provides significant traffic sensing accuracy. We also describe the lessons learned from our deployment and inspiring observations.

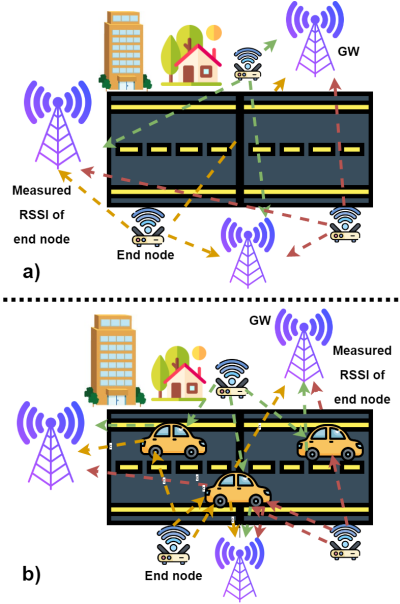

**Figure 2: An illustration of the proposed method for traffic sensing based on LoRa.**

## 3 PROPOSED METHOD

Figure 2 illustrates the proposed method and the LoRaWAN deployment of end nodes in urban areas to perform traffic sensing. The authors use RSSI measurements from multiple LoRaWAN nodes to estimate the proximity of automobile traffic.

### 3.1 Environment Coverage Effect on RSSI

By increasing the popularity of LoRaWAN, the number of implemented LoRaWAN end nodes in different environments is increasing. The end nodes are installed in fixed points and transmit packets to a server periodically. Gateways receive the packets and measure their RSSI. The actual value of signal path loss from sender to receiver can be calculated by using measure RSSI throw Equation 1.

$$PL = RSSI + SNR + P_{tx} + G_{tx} \qquad (1)$$

where $RSSI$, $SNR$, $P_{tx}$, and $G_{tx}$ represent the measured RSSI and SNR in the gateway, the transmission power, and the gain of the transmitter, respectively. Since the distance of the end node to the gateway is fixed, the expected $PL$ can be conveniently modeled using the Log-distance path loss (LDPL) model calculated in Equation (2).

$$PL(d) = PL(d_0) + 10 n log(d/d_0) + X\sigma \qquad (2)$$

where $n$ is the path loss exponent, $PL(d_0)$ is the path loss value at a reference distance from the receiver, $d$ is the distance between an end node and the LoRaWAN gateway, $d_0$ is the reference distance (1 m), and $X\sigma$ is a zero-mean Gaussian random variable to account for the shadow fading.

We assume the most important reason for the difference between the measured PL and EPL of the end node, implemented in the fixed point is due to attenuation caused by the dynamic of the environment. In this paper, attempts are made to predict environmental traffic congestion by computing the difference between the measured PL and EPL. For this purpose, we use a Support Vector Machine (SVM) to get the measured RSSI of the end node from the gateway and use them as features to provide a classifier that estimates the traffic of the environment.

### 3.2 Support Vector Machine Classification

There are several works that survey and apply ML approaches to wireless communication in different domains [15, 20, 23–34, 38? –40]. that are applied to The SVM algorithm was introduced in the late 1970s by Vapnik and his group. It is one of the most popular and widely used supervised and kernel-based learning algorithms among machine learning algorithms. In particular, SVM identifies an optimal hyperplane to divide the input data into a discrete number of predefined classes using the training data. In the proposed method, we use SVM to map the measured RSSI of end nodes to the amount of traffic in the environment. To do so, different classes are defined based on the amount of traffic in the environment. The measured RSSIs are used as features for classification. An illustration of the proposed method is shown in Fig ??. As shown in Fig 2 (a), in a stable environment, the propagation model of the signal is almost constant. By increasing the traffic and the dynamics of the environment, the propagation model of the signal has changed (Fig 2 (b)). Variation of the signal propagation model is the reason for changes in the measured RSSI and path loss parameters. In the proposed method, different classes of proposed traffic sensing approach are labeled based on the number of cars (this approach can be extended to other objects such as humans) in the environment, and the reported RSSI of the end node by different gateways. As mentioned, SVM is used as a classifier to provide an estimation of the traffic class of the environment based on the measured RSSI.

## 4 DEPLOYMENT AND EVALUATION

In this section, first, we present the components and the architecture of the implemented LoRaWAN testbed, which consists of three deployed LoRaWAN gateways (on the rooftop of a central building) and one end node installed in the middle of the parking lot. Then, important deployment planning factors are presented.

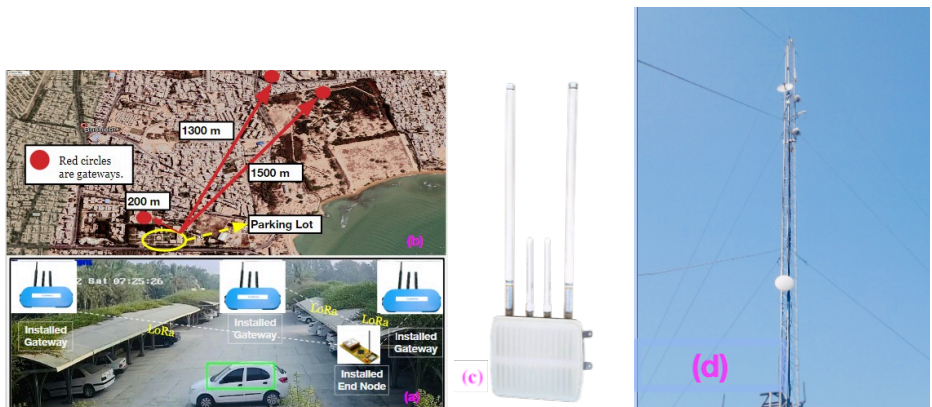

**Figure 3: Experimental Setup. From left to right: (a) Experimental setup in a parking lot where measurements were collected from 16 different end node positions. (b) The bird's eye view (taken by Google Maps) of the entire area identifies the location of the gateways, their precise distances where they are installed, and the location of the parking lot. (c) LoRaWAN gateway that is assembled with Lora Radio and LoRaWAN Antennas (d) An example of a tower where LoRaWAN gateways are installed (In total, three gateways are installed on three towers shown in (d)).**

## 4.1 LoRaWAN Experimental Testbed

The hardware used and the testbed organization are presented and discussed in this Section. The measurement was conducted in a typical parking environment with a size equal to 100 m by 35 m in the university. The core components of this implemented testbed are three LoRaWAN gateways and LoRaWAN end nodes, and their altering positions. Each gateway is equipped with an MCU and an SX1301 digital baseband transceiver has been installed on one of the three antenna towers with a height between 30 and 35 m. LoRaWAN end node is implemented and installed using an MCU, a transceiver SX1276 mounted on the roof of parking with a height equal to 3 m. The LoRaWAN end node transmits packets to the gateways at intervals equal to 60 seconds. The time between sensing the environment and receiving the measurements in a server or database plays an important role in the proposed method approach. Data capture at higher rates provides more detailed information about the changes in the environment. Latencies on the order of seconds or minutes enable near-real-time applications. All the packets are transmitted with a spreading factor, bandwidth, coding rate, and channel equal to 7, 125 kHz, 4/5, and 868 kHz, respectively. We logged over 7000 records sensed by at least two gateways in total for weeks.

## 4.2 Deployment Planning

For detecting the best places for placement of end nodes in the parking lot, a fingerprint radio map of the testbed containing RSSI data for 28 points was measured to characterize signal function. Then, 16 points based on different RSSI signal levels were selected out of 28 points. In the selection process, points were selected such that the variance of the changes in the value of the point measured RSSI by the various gateways is significant compared to other points.

Next, according to the size of the parking lot and its maximum capacity (= 50 cars), five different classes were considered indicating five different traffic levels. These classes include Class 1 (up to 17 cars), Class 2 (between 18 and 24 cars), Class 3 (between 25 and 31 cars), Class 4 (between 32 and 38 cars), and Class 5 (more than 39 cars). Such a classification could also be applied as a universal approach for traffic sensing in any parking lot depending on its size. It is worth mentioning that, the size and density of the deployments will depend on the environment size, the number of objectives, and the characteristics of the measured quantities. in total for weeks.

## 5 RESULTS

Our experiments serve two primary thrusts. First, we measure the basic performance of our RSSI in terms of traffic sensing. We are then interested to see how the dynamicity of the environment impacts the accuracy of our proposed solution and how we can recover from sudden changes in the environment. To implement the SVM algorithms, we use the MATLAB simulator and to make a reliable evaluation, cross-validation was applied during the simulation. Then, the average distance prediction error is reported. To perform the evaluation, we split observations of the end node by 70% for training and 30% for the detection test. For evaluation of the proposed traffic sensing method, four evaluation parameters of Accuracy, Precision, Recall, and False Detection Rate are computed. These parameters are presented in the following.

- **Accuracy**: fraction of correct predictions with regard to the model under evaluation, cf. Eq. 3,

$$\text{Accuracy} = \frac{TP + TN}{TP + TN + FP + FN} \quad (3)$$

where $TN$ are True Negatives, $TP$ are True Positives, $FP$ are False Positives, and $FN$ are False Negatives.

- **Precision**: represents the qualities of accurate detection to all detection, cf. Eq. 4.

$$\text{Precision} = \frac{TP}{TP + FP} \quad (4)$$

- **Recall**: represents the fraction of instances that were classified correctly, cf. Eq. 5,

$$\text{Recall} = \frac{TP}{TP + FN} \quad (5)$$

- **False Detection Rate**: calculated using Equation 6.

$$\text{False Detection Rate} = \frac{FP}{FP + TN} \quad (6)$$

The simulation results of SVM for the different considered classes of the dataset are shown in Figure 4. As it is clear in the Figure, the proposed method provides an acceptable diagnosis of the amount of traffic congestion in the environment by predicting the class of environment.

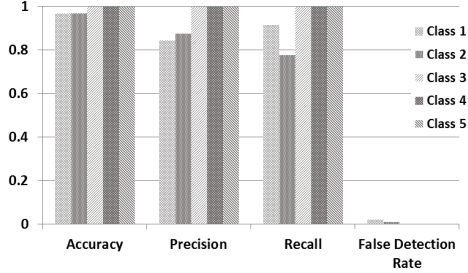

**Figure 4: Evaluation factors.**

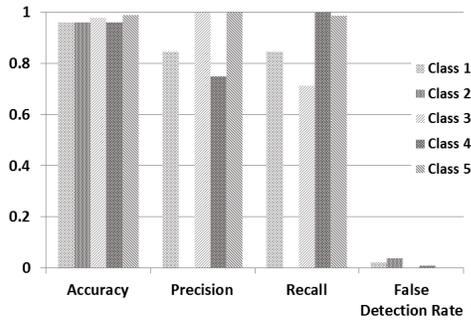

**Figure 5: Effect of end node position in accuracy.**

## Curiosities & Observations

***Lessons Learned: Clustering VS classification***. To evaluate the proposed method, K_means as a clustering method is simulated too. The k-means function partitions data into k mutually exclusive clusters and returns the index of the cluster to which it assigns each observation. The accuracy result of K-means clustering algorithms is presented in Table 1. As it is clear in the Table, k-means has low accuracy in predicting the traffic congestion level of the environment.

**Table 1: K-means clustering algorithms accuracy.**

|  | Class 1 | Class 2 | Class 3 | Class 4 | Class 5 |
|---|---|---|---|---|---|
| Accuracy | 0.72 | 0.84 | 0.68 | 0.52 | 0.61 |

***Why does SVM yield better results?*** RSSI is very vulnerable to noise and interference. As the proposed model is using RSSI for estimation, the measured RSSI as input of the adopted ML method can have noise. These noisy data in unsupervised data can lead to a higher error rate. So, supervised learning methods lead to better estimation and SVM is one of the best-supervised learning algorithms.

***Post Deployment Assessment***. In the deployment process, 16 different points were specified for end-node deployment. During the evaluation of the proposed method by using measured RSSI of each one of them, we figure out that the place of the end node has an important effect on the accuracy of traffic-level prediction. Based on the result, it is important to implement the end node in the middle of the target area to have an acceptable experience of environmental changes. By having an acceptable experience of the environment, a more accurate estimation of the traffic of the environment can be achieved. The achieved estimation accuracy result of an implemented end node in the beginning point of the parking is shown in Fig 5.

## 6   FUTURE WORK & CONCLUSION

In this study, we explored the usage of LoRaWAN end nodes as traffic sensing elements to offer a practical traffic control solution. The monitored Received Signal Strength Indicator (RSSI) factor is reported to the gateways to assess the traffic congestion of the environment. Our technique utilizes LoRaWAN as a long-range communication technology to provide a large-scale system. This paper detailed our experiences with the design and real implementation of this system across a harbor area that stretches for miles. Our proposed traffic sensing approach is a first step in utilizing LoRaWAN's advantages for traffic management and traffic monitoring in urban areas where the environment's dynamic is a massive challenge. The result of this study call for future research to strengthen the measurement accuracy and scalability of our solution for different cities with different architecture and sizes.